\documentclass[letterpaper, 10 pt, conference]{ieeeconf}  % Comment this line out if you need a4paper

\IEEEoverridecommandlockouts                              % This command is only needed if 
\IEEEoverridecommandlockouts                              % This command is only needed if 
\let\labelindent\relax
\usepackage{graphics} % for pdf, bitmapped graphics files

\usepackage{xcolor}
\usepackage{framed,enumitem} 
\usepackage{epsfig} % for postscript graphics files
\usepackage{times} % assumes new font selection scheme installed
\usepackage{amsmath, bm}
\usepackage{amssymb}  % assumes amsmath package installed
\usepackage{arydshln}
\usepackage{mathrsfs}
\usepackage{amsfonts}
\usepackage{algorithm}
\usepackage[noend]{algpseudocode}
\usepackage[noadjust]{cite}
\usepackage{stfloats}
\usepackage{bm}
\usepackage{algorithm}
\usepackage[noend]{algpseudocode}
\newcommand{\YB}[1]{{\color{black}#1}}

\newcommand{\TR}[1]{{\color{black}#1}}

\title{\LARGE \bf
\YB{Frequency Domain Auto-tuning of Structured LPV Controllers for High-Precision Motion Control*}
}

\author{Yorick Broens, Hans Butler and Roland T\' oth% <-this % stops a space
\thanks{*This work has received funding from the ECSEL Joint Undertaking (JU) \YB{programme} under grant agreement No 875999 and from the European Union within the framework of the National Laboratory for Autonomous Systems (RRF-2.3.1-21.2022-00002).}% <-this % stops a space
\thanks{\YB{Y. Broens}, \YB{H. Butler} and \YB{R. T\'oth} are with the Department of Electrical Engineering, Eindhoven University of Technology, Eindhoven, The Netherlands. \YB{H. Butler} is also affiliated with ASML, Veldhoven, The Netherlands. \YB{R. T\'oth} is also affiliated with the Systems and Control Laboratory, HUN-REN Institute for Computer Science and Control, Hungary,  ({\tt  email: Y.L.C.Broens@tue.nl}). }
}

\begin{document}

\maketitle
\thispagestyle{empty}
\pagestyle{empty}

%%%%%%%%%%%%%%%%%%%%%%%%%%%%%%%%%%%%%%%%%%%%%%%%%%%%%%%%%%%%%%%%%%%%%%%%%%%%%%%%
\begin{abstract}
Motion systems are a vital part of many industrial processes. However, meeting the increasingly stringent demands of these systems, especially concerning precision and throughput, requires novel control design methods that can go beyond the capabilities of traditional solutions. Traditional control methods often struggle with the complexity and position-dependent effects inherent in modern motion systems, leading to compromises in performance and a laborious task of controller design. 
This paper addresses these challenges by introducing a novel structured feedback control auto-tuning approach for multiple-input multiple-output (MIMO) motion systems. By leveraging frequency response function (FRF) estimates and the linear-parameter-varying (LPV) control framework, the proposed approach automates the controller design, while providing local stability and performance guarantees. Key innovations include norm-based magnitude optimization of the sensitivity functions, an automated stability check through a novel extended factorized Nyquist criterion, a modular structured MIMO LPV controller parameterization, and a controller discretization approach which preserves the continuous-time (CT) controller parameterization. The proposed approach is validated through experiments using a state-of-the-art moving-magnet planar actuator prototype.

%In the realm of technological advancements, motion systems stand as a vital frontier, enhancing the performance and reliability of various industrial processes. However, meeting the increasingly stringent demands of industrial processes, especially concerning precision and throughput, necessitates sophisticated control designs. Traditional control methods often struggle with the complexity and position-dependent effects inherent in modern motion systems, leading to compromises in performance. This paper addresses these challenges by introducing a novel structured feedback control approach for Multiple-Input Multiple-Output (MIMO) systems. Leveraging Frequency Response Function (FRF) models and the Linear Parameter Varying (LPV) control framework, our approach automates controller design while ensuring local stability and performance guarantees. Key innovations include norm-based magnitude optimization, extended factorized Nyquist checks, and a modular controller design based on Linear Fractional Representations (LFRs). The contributions of this paper lie in the development of a structured LPV MIMO feedback control auto-tuning parameterization, a novel local stability check, and a structured MIMO feedback control auto-tuning approach capable of direct discrete-time controller synthesis. These contributions pave the way for improved control design in modern motion systems, preserving both performance and modularity of the controller design.
\end{abstract} 

%%%%%%%%%%%%%%%%%%%%%%%%%%%%%%%%%%%%%%%%%%%%%%%%%%%%%%%%%%%%%%%%%%%%%%%%%%%%%%%%
\vspace{-.1cm}
\section{Introduction}
\label{Section:Introduction}
\vspace{-.1cm}

In recent years, there has been a significant focus on motion systems. These systems play a crucial role in enhancing the performance and reliability of various industrial processes and manufacturing systems, such as wafer scanners, industrial printers, pick-and-place machines and wire bonders, see \cite{Butler,ohnishi1996motion,bukkems2006piecewise,qu2016motion,6225187}. Traditionally, motion control design for MIMO systems has leaned heavily on superior mechanical design principles, emphasizing factors such as high stiffness and reproducibility, see \cite{Oomen}. This approach has led to motion dynamics primarily characterized by \emph{rigid-body} (RB) dynamics, facilitating the use of RB decoupling strategies to effectively mitigate low-frequent channel interaction%inherent in MIMO systems
, see \cite{Steinbuch2013}. 
In industrial settings, the integration of RB decoupling with \emph{sequential loop-closing} (SLC) controller design is a common practice due to several practical advantages. Primarily, SLC facilitates the application of established loop-shaping techniques, see \cite{STEINBUCH1998278}. Moreover, it simplifies motion control design by relying on non-parametric models, i.e., FRFs, thereby obviating the necessity for precise parametric system identification. Despite its advantages, the process of motion control design via SLC presents formidable challenges, particularly for high number of inputs and varying dynamics.

Modern motion systems frequently experience position-dependent effects, as noted in \cite{Butler}. To address these effects, it becomes necessary to adapt the RB decoupling technique to account for positional variations, thereby enabling the implementation of SLC-based controller designs. While this approach is effective in achieving RB decoupling, it often leads to persistent high-frequency position-dependent couplings, which compel the SLC-based design to prioritize robustness, ultimately resulting in a degradation of performance.

To automate motion control design in practice using only FRF data, various auto-tuning methods have been developed for single-input single-output (SISO) systems, as documented in \cite{zhu2006simple,van2018frequency,hwang2002solution,panagopoulos1999design}. However, these methods are limited in their ability to accommodate MIMO systems and position-dependent dynamics. Despite this, a LPV MIMO auto-tuning approach that utilizes FRFs exclusively is presented in \cite{9667108}. Nonetheless, the application of this approach to structured controller synthesis is challenging due to the orthogonal basis function parameterization of the feedback controller. Alternatively, optimal gain controller synthesis techniques, such as robust $\mathcal{H}_\infty$ and LPV $\mathcal{L}_2$ control, have emerged to provide efficient control design for MIMO dynamics even with position-dependent characteristics. Also, efficient tools for structured controller synthesis based on these approaches, i.e., using $\mathrm{Hinfstruct}$, see \cite{gahinet2011decentralized}, have been introduced. Nevertheless, the deployment of these latter approaches necessitate precise low-order parametric models capable of accurately capturing the high-frequency position-dependent channel interactions, which poses a formidable challenge in the context of modern system identification. 
%To automate motion control design in practise using FRF data only, various auto-tuning methods have been developed for SISO systems, see \cite{{zhu2006simple,van2018frequency,hwang2002solution,panagopoulos1999design}
%}. However, these approaches exhibit limitations in accommodating MIMO and position dependent dynamics. Alternatively, optimal gain controller synthesis techniques, such as robust $\mathcal{H}_\infty$ and LPV $\mathcal{L}_2$ control, have emerged to provide efficient control design for MIMO dynamics even with position-dependent characteristics. Also, efficient tools for structured controller synthesis based on these approaches, i.e., using $\mathrm{Hinfstruct}$, see \cite{gahinet2011decentralized}, have been introduced. Nevertheless, the deployment of these latter approaches necessitate precise low-order parametric models capable of accurately capturing the high-frequency position-dependent channel interactions, which poses a formidable challenge in the context of modern system identification. 

%To overcome the gap between existing SISO approaches and required MIMO position dependent capabilities, this paper presents a novel frequency domain-based auto-tuning approach for LPV MIMO systems, relying solely on FRFs of the motion system, 
%thus bypassing the need for complex parametric identification while providing local stability and performance guarantees.

To facilitate the synthesis of structured LPV MIMO feedback controllers, this paper presents a novel frequency domain-based auto-tuning approach for LPV MIMO systems, relying solely on FRFs of the motion system, 
thus bypassing the need for complex parametric identification while providing local stability and performance guarantees.
%In the extension of the current LTI auto-tuning schemes, 
The main contributions of this paper are:
\begin{itemize}
\item[(C1)] Development of a novel structured LPV MIMO feedback controller parameterization for auto-tuning, ensuring the modularity of controller design.
\item[(C2)] Development of a MIMO stability check for diagonal and full block controllers, relying only on FRF data.
\item [(C3)] Development of a novel discrete-time LPV controller implementation, preserving the CT parameterization.
\end{itemize}

This paper is organized as follows. Section \ref{SectionII_ProblemFormulation} presents the problem formulation, followed by the proposed structured feedback control parameterization in Section \ref{SectionIII_Approach}. In Section \ref{SectionIV_Optimization}, the optimization problem is introduced, encompassing an automated stability check, closed-loop performance shaping and a novel controller implementation approach. Section \ref{SectionV_ExperimentalValidation} presents experimental results of the frequency domain-based LPV MIMO structured feedback control auto-tuner on a state-of-the-art \emph{moving-magnet planar actuator} (MMPA) prototype, while Section \ref{SectionVI_Conclusion} draws conclusions on the presented work.

\vspace{-.2cm}
\section{Problem Formulation}
\label{SectionII_ProblemFormulation}
\vspace{-.1cm}
\subsection{Background}
\vspace{-.1cm}

\begin{figure}[t]
    \centering
    \includegraphics[trim={2.9cm 0cm 3.4cm 1.4cm},clip,width=\linewidth,height = 5.5cm]{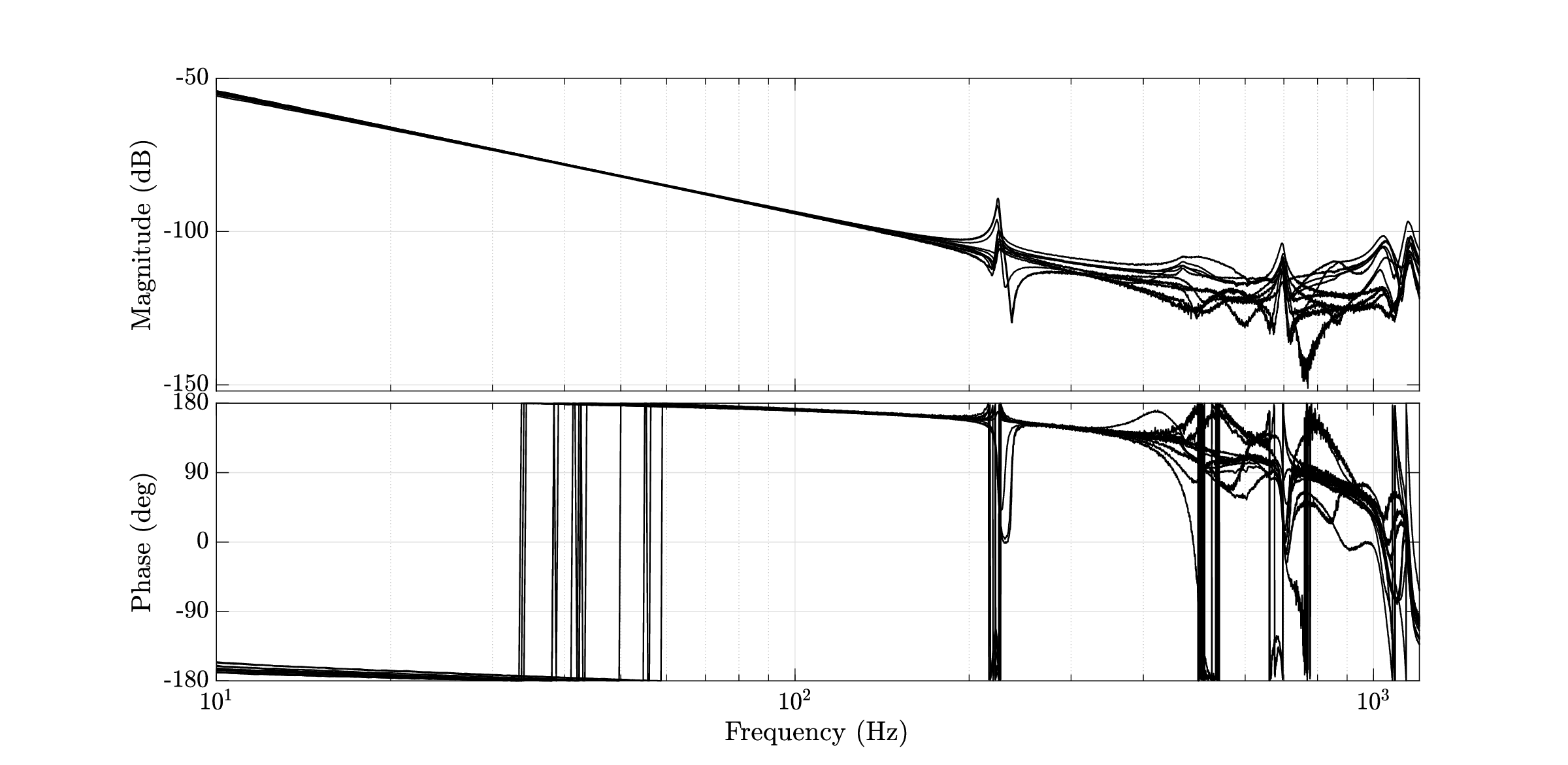}
    \vspace{-.8cm}
    \caption{Set of local FRFs of a high-precision moving-magnet planar actuator prototype, illustrating the
    high-frequency position dependent flexible dynamics for the RB decoupled transfer in $R_y$-direction.}
    \label{fig:SectionII_FigureI}
    \vspace{-.4cm}
\end{figure}

Many modern motion systems exhibit position dependent effects due to their increasingly complex nature, see Figure \ref{fig:SectionII_FigureI}.  
A common cause for this is the relative actuation and sensing of the moving-body, necessitating position dependent RB coordinate frame transformations to establish a relationship between the point of interest on the moving-body, the actuation forces and the actual position measurements.
To accurately capture these effects, such systems are often represented in LPV form, where the position dependency is encapsulated within a scheduling vector, see \cite{5714737}. Consider the equations of motion of a motion system that exhibits position dependent effects in the input and output:
\vspace{-.1cm}
\begin{equation}
    M\ddot{x}(t) + D\dot{x}(t) + Kx(t) = G(p(t))f(t),
\label{SectionII_PhysicalEquationsofMotion} 
\end{equation}

\vspace{-.1cm}
\noindent where $M, D$ and $K$  $\in \mathbb{R}^{n_x \times n_x}$ are the real symmetric mass, damping, and stiffness matrices and $G(p(t)) \in \mathbb{R}^{n_x\times n_f}$ maps the forces acting on the moving body, $f(t)$, to its center of gravity based on the scheduling vector $p(t):\mathbb{R}\rightarrow \mathbb{P}\subseteq\mathbb{R}^{n_p}$. To allow for independent control of the mechanical degrees of freedom, (\ref{SectionII_PhysicalEquationsofMotion}) is typically represented in modal form, see \cite{gawronski2004dynamics}. This is achieved through a state transformation $x(t)=\tilde{V}\eta(t)$, where $\tilde{V}=M^{-\frac{1}{2}}V$. The eigenvector matrix $V$ is derived from the characteristic dynamical equation $KV=MV\Lambda$, where $\Lambda$ is the eigenvalue matrix. Grouping of the states per mode is achieved through a secondary state transformation $\eta(t) = T(\eta_{_{\mathrm{RB}}}^\top (t)\ \eta_{_{\mathrm{FM}}}^\top (t))^\top$ with:
\vspace{-.1cm}
\begin{equation}
    T=\left(I_{n_x \times n_x}\otimes (1 \ \ 0)^\top \ \ I_{n_x \times n_x}\otimes (0 \ \ 1)^\top \right),
\end{equation}

\vspace{-.1cm}
\noindent where $\otimes$ corresponds to the \emph{Kronecker-product}. Furthermore, the corresponding partitioned modal state-space representation of the motion system, denoted by $\mathcal{P}$, corresponds to:
\begin{equation}
\resizebox{.91\hsize}{!}{$
        \left ( \begin{array}{c}
             \dot{\eta}_{_{\mathrm{RB}}}(t) \\ \dot{\eta}_{_{\mathrm{FM}}}(t)  \\ \hdashline y(t)
        \end{array}\right) =
         \left ( \begin{array}{cc:c}
             A_{_{\mathrm{RB}}}  & 0 &B_{_{\mathrm{RB}}}(p(t))  \\  0& A_{_{\mathrm{FM}}}  & B_{_{\mathrm{FM}}}(p(t))  \\ 
             \hdashline 
             C_{_{\mathrm{RB}}}(p(t))   &C_{_{\mathrm{FM}}}(p(t))  & 0
        \end{array}\right) \left ( \begin{array}{c}
             \eta_{_{\mathrm{RB}}}(t) \\ \eta_{_{\mathrm{FM}}}(t) \\ \hdashline f(t)
        \end{array}\right),$}
    \label{SectionII_ModalDynamics}
\end{equation}

\noindent where $\left(\cdot \right)_{_{\mathrm{RB}}}$ are the system matrices that correspond to the rigid body modes and $\left(\cdot \right)_{_{\mathrm{FM}}}$ are the system matrices that coincide with the flexible modes. In the industry, motion control design for these type of systems is simplified through position dependent RB decoupling, which, in this case, is achieved by introducing the input and output decoupling matrices:
    \begin{subequations}
    \begin{align}
    T_\mathrm{u}&= \left((I_{n_{_{\mathrm{RB}}} \times n_{_{\mathrm{RB}}}} \otimes \begin{bmatrix}0 & 1 \end{bmatrix} )B_{_{\mathrm{RB}}}(p(t))\right)^\dagger, \\
    T_\mathrm{y}&= \left(C_{_{\mathrm{RB}}}(p(t))(I_{n_{_{\mathrm{RB}}} \times n_{_{\mathrm{RB}}}} \otimes \begin{bmatrix}1 & 0 \end{bmatrix} )^\top \right)^\dagger,
    \label{Section2_Equation_6}
\end{align}
\end{subequations}

\noindent where $n_{_{\mathrm{RB}}}$ corresponds to the number of RB modes of the system. In this context, the RB decoupled system is given by $\tilde{\mathcal{P}} = T_\mathrm{y}{\mathcal{P}}T_\mathrm{u}$. It is noteworthy to observe that introduction of the decoupling matrices results in an elimination of the position dependency in the RB dynamics, see Figure \ref{fig:SectionII_FigureI}. Nonetheless, position dependent interaction still persists due to the flexible dynamics, necessitating robustified controller design at the cost of performance. An additional important observation is that in case the scheduling vector is constant, i.e., $p(t) = \boldsymbol{\mathrm{p}}$ for all $t \in \mathbb{R}$, $\Tilde{\mathcal{P}}$ becomes an LTI system, which is often referred to as \emph{local} or \emph{frozen dynamics} of the LPV system. For a given fixed $\boldsymbol{\mathrm{p}}$, the \emph{Fourier transform} of the local signal relation is given by:
\begin{equation}
    Y(j\omega) = \tilde{\mathcal{P}}_{\boldsymbol{\mathrm{p}}}(j\omega) U(j\omega),
    \label{SectionII_FourierTransform}
\end{equation}

\vspace{-.1cm}

\noindent where $j$ is the imaginary unit, $\omega \in \mathbb{R}$ corresponds to the frequency and $\tilde{\mathcal{P}}_{\boldsymbol{\mathrm{p}}}(j\omega)$ denotes the \emph{local frequency response function} (lFRF) of $\Tilde{\mathcal{P}}$. In this context, a set of lFRFs, denoted by $\lbrace \tilde{\mathcal{P}}_i \rbrace_{i=1}^{n}$, is obtained through closed-loop identification approaches for various forced equilibria of the system, i.e., around various operating points $\boldsymbol{\mathrm{p}}$, thereby capturing the complex position dependent high-frequent effects in an accurate manner.
Moreover, this set of lFRFs can be used for analyzing the local performance of a designated structured controller ${\mathcal{K}} \in \mathbb{R}^{n_{_{\mathrm{RB}}} \times n_{_{\mathrm{RB}}}}$ through the assessment of the magnitude constraints associated with closed-loop sensitivities across various frequency ranges. In this context, one may contemplate an optimization problem aimed at synthesizing structured controllers given a weighted plant. Similar to optimal gain-based control design, see \cite{skogestad2007multivariable}, the construction of such a weighted plant is facilitated by shaping performance channels through frequency dependent filters.% $w_1, w_2, z_1, z_2 \in \mathbb{R}^{n_\mathrm{RB}}$.

\vspace{-.1cm}
\subsection{Problem Statement}
\vspace{-.1cm}

The problem that is being addressed in this paper is to develop a frequency-domain structured LPV MIMO feedback control auto-tuning approach by using lFRF measurements. We aim to accomplish this under the following requirements:
\begin{itemize}
    \item [(R1)] The system is locally stabilized by ${\mathcal{K}}$ for all $ \tt p \in \mathbb{P}$.
    \item [(R2)] The control synthesis solely relies on lFRFs of the system, thus avoiding complex parametric identification.
\end{itemize}

%, see Figure \ref{fig:RigidBodyAutotunerConfiguration}. 

%To exploit the advantages offered by the RB decoupling of $\tilde{\mathcal{P}}$, we propose a structured controller $\tilde{\mathcal{K}}$ that is partitioned into two distinct components:  (i) a low-frequency LTI controller, directed towards shaping the RB dynamics to achieve desired characteristics, and (ii) a high-frequency LPV controller, aimed at actively addressing position-dependent flexible dynamics. To allow for modular control design, we propose a novel structured parameterization approach for LPV MIMO controllers, capitalizing on the structure of \emph{linear fractional representations} (LFRs). This parameterization approach involves the extraction of controller parameters within an upper diagonal interconnection, see Figure \ref{fig:RigidBodyAutotunerConfiguration}, facilitating a modular approach to the MIMO controller design.

\section{Controller Parameterization}
\label{SectionIII_Approach}
%In this Section, we propose a novel modular LPV MIMO structured feedback control parameterization for auto-tuning. The presented approach exploits the linear fractional representation of individual  first and second order filters to allow for modular controller design of the structured LPV MIMO feedback controller. This is achieved by extracting the controller parameters into a lower block diagonal forming a controller, while the resulting disconnected (LTI) dynamic interconnection is collapsed in the generalized plant, i.e., $\mathcal{K}_{\mathrm{LTI}}$, $\mathcal{K}_{\mathrm{LPV}}$ in Figure \ref{fig:RigidBodyAutotunerConfiguration}.
%In this Section, a novel modular LPV MIMO structured feedback control parameterization is presented for auto-tuning purposes. Consider the feedback control partitioning illustrated in Figure \ref{fig:RigidBodyAutotunerConfiguration}, where the RB decoupling of the system is exploited by partitioning the controller into a low-frequency LTI controller and a high-frequency LPV controller. The control of RB modes is typically achieved through a PID type of controller, see \cite{skogestad2007multivariable}, where the time-domain representation of a PI controller corresponds to:
This section introduces a novel modular LPV MIMO structured feedback controller parameterization for auto-tuning. To exploit the characteristics of typical motion systems, the structured feedback controller $\mathcal{K}$ is divided into two main components as illustrated in Figure \ref{fig:SectionIII_ControllerStructure}:
\begin{itemize}
    \item [(i)] A low-frequency LTI controller, aiming at shaping the RB dynamics to achieve the desired characteristics.
    \item [(ii)] A high-frequency LPV controller, designed to address position-dependent flexible dynamics.
\end{itemize}

\noindent 
To ensure modularity and scalability in structured controller design, a novel structured parameterization approach is proposed using \emph{linear fractional representations} (LFRs). This approach extracts the controller parameters into upper-diagonal interconnections, resulting in diagonal parameter matrices that can be optimized based on performance specifications. The remaining controller dynamics are incorporated into the generalized plant for optimization.
%Consequently, in the resulting LFR description of the controller, i.e., ${C}_{\mathrm{LPV}}$ and ${C}_{\mathrm{LTI}}$, exclusively comprises integrator connections. 
%Consequently, the resulting generalized plant description, i.e., $\mathcal{C}_{\mathrm{PI}}, \mathcal{C}_{\mathrm{LF}}$ and $\mathcal{C}_{\mathrm{N}}$, exclusively comprises integrator connections.
\subsection{Low-frequency LTI controller design}

The control of RB modes conventionally employs a PID-type controller, see \cite{skogestad2007multivariable}, which is constructed by combining a PI-type controller with lead filters in a cascaded manner. It is worth noting that this controller configuration can also be implemented using a parallel controller structure. Nonetheless, the proposed controller parameterization accommodates both architectures as will be discussed in Subsection \ref{SectionIII_SubsectionC}. Consider the time-domain representation of a PI-controller:
\vspace{-.1cm}
\begin{subequations}
    \begin{align}
    \dot{x}_{_{\mathrm{PI}}}(t) &=  u_{_{\mathrm{PI}}}(t), \\
    y_{_{\mathrm{PI}}}(t) &= K_p x_{_{\mathrm{PI}}}(t),
    \label{PICTRON}
\end{align}
\end{subequations}

\vspace{-.05cm}
\noindent where $u_{_{\mathrm{PI}}}(t),y_{_{\mathrm{PI}}}(t) \in \mathbb{R}^{n_{_{\mathrm{RB}}}}$ correspond to the input and the output of the PI-controller and $K_p \in \mathbb{R}^{n_{_{\mathrm{RB}}}\times n_{_{\mathrm{RB}}}}$ is a diagonal 
matrix containing the proportional gains, ensuring that the local loop transfers, i.e., $\lbrace \tilde{\mathcal{P}}_i{\mathcal{K}}_i\rbrace_{i=1}^n $, cross the 0 dB line at the desired target bandwidths. To ensure closed-loop stability of the system, lead filters are often integrated alongside PI-controllers in a cascaded manner. The time-domain representation of a first order lead-filter is given by:
\vspace{-.5cm}
\begin{subequations}
    \begin{align}
    \dot{x}_{_{\mathrm{LF}}}(t) &=  -\Omega_2x_{_{\mathrm{LF}}}(t)+ u_{_{\mathrm{LF}}}(t), \\
    y_{_{\mathrm{LF}}}(t) &= (\Omega_1-\Omega_2) x_{_{\mathrm{LF}}}(t)+ u_{_{\mathrm{LF}}}(t), 
    \label{LFFILT}
\end{align}
\end{subequations}

\vspace{-.05cm}
\noindent where $u_{_{\mathrm{LF}}}(t),y_{_{\mathrm{LF}}}(t) \in \mathbb{R}^{n_{_{\mathrm{RB}}}}$ correspond to the input and the output of the lead filter, and, $\Omega_1 ,\Omega_2 \in \mathbb{R}^{n_{_{\mathrm{RB}}}\times n_{_{\mathrm{RB}}}}$ are diagonal matrices containing the cut-off frequencies of the differentiatiors and integrators. To accommodate sufficient phase lead at the target bandwidth, i.e., $30\leq \phi \leq 60$ degrees phase margin while being subject to the integral action of the PI controller, a third order lead filter is typically required.

\subsection{High-frequency LPV controller design}

To actively combat position dependent flexible dynamics, position dependent notch filters are employed. Consider the time-domain representation of a LPV notch filter:
\begin{subequations}
\begin{align}
        \dot{x}_{_{\mathrm{N}}}(t) &=  A_{_{\mathrm{N}}}(p(t)) 
 x_{_{\mathrm{N}}}(t)+ B_{_{\mathrm{N}}}u_{_{\mathrm{N}}}(t), \\
    y_{_{\mathrm{N}}}(t) &= C_{_{\mathrm{N}}}(p(t)) x_{_{\mathrm{N}}}(t)+  D_{_{\mathrm{N}}} u_{_{\mathrm{N}}}(t)  \ \ ,
    \label{NotchFilter}
\end{align}
    \end{subequations}

\noindent where the state-space matrices are defined as follows:
\begin{subequations} 
\begin{align}
        A_{_{\mathrm{N}}}(p(t)) &= \begin{pmatrix}
             -2\beta_2(p(t))\omega_2(p(t))& -\omega_2^2(p(t))  \\
             I_{n_{_{\mathrm{RB}}}\times n_{_{\mathrm{RB}}}}&0_{n_{_{\mathrm{RB}}}\times n_{_{\mathrm{RB}}}} 
       \end{pmatrix}, \\
        B_\mathrm{N} &= \begin{pmatrix}
             I_{n_{_{\mathrm{RB}}}\times n_{_{\mathrm{RB}}}}&0_{n_{_{\mathrm{RB}}}\times n_{_{\mathrm{RB}}}}
       \end{pmatrix}^\top ,\\
          C_{_{\mathrm{N}}}(p(t)) &=    \resizebox{0.65\hsize}{!}{$
 \begin{pmatrix}
 2(\beta_1(p(t))\omega_1(p(t))-\beta_2(p(t))\omega_2(p(t))) \\\omega_1^2(p(t))-\omega_2^2(p(t))
        \end{pmatrix} ^\top $},
        \\
        D_{_{\mathrm{N}}} &= I_{n_{_{\mathrm{RB}}}\times n_{_{\mathrm{RB}}}}.
        \label{posdepnotchfilter}
\end{align}
    \end{subequations}

\noindent Here, $\beta_1(p(t)), \beta_2(p(t)) \in \mathbb{R}^{n_{_{\mathrm{RB}}} \times n_{_{\mathrm{RB}}}}$ represent diagonal matrices containing the damping ratios of the notch filter, which regulate the amplitude suppression at given notch frequencies $\omega_1(p(t)), \omega_2(p(t)) \in \mathbb{R}^{n_{_{\mathrm{RB}}} \times n_{_{\mathrm{RB}}}}$.

\begin{figure}[t]
    \centering
    \includegraphics[width=\linewidth]{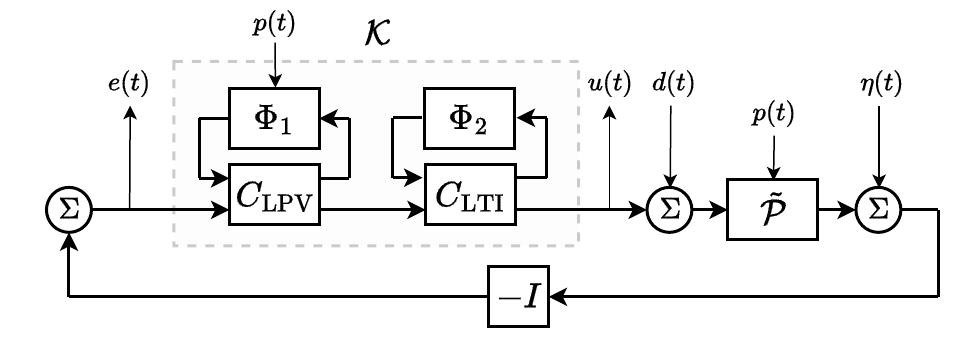}
    \vspace{-.6cm}
    \caption{Closed-loop motion control interconnection, where $\mathcal{K}$ is partitioned into a low-frequency LTI controller and a high-frequency LPV controller.}
    \label{fig:SectionIII_ControllerStructure}
    \vspace{-.4cm}
\end{figure}

\subsection{Controller interconnection}
\label{SectionIII_SubsectionC}

%\begin{figure}[b]
    %\centering
    %\includegraphics[width=.7\linewidth%]{Figures/PKFORMCONTROLLER.pdf}
%    \caption{Caption}
%    \label{fig:enter-label}
%\end{figure}

To allow for modularity and scalability of the controller design, the filters presented in (6), (7) and (8) are reformulated in LFR form as illustrated in Figure \ref{fig:SectionIII_ControllerStructure}. This is achieved by extracting the controller parameters into an upper diagonal interconnection, i.e., $\tilde{u}_i = \Phi_i \tilde{y}_i$, yielding the LFR:
\begin{equation}
    \left( \begin{array}{c}
         \tilde{y}_i(t) \\  {y}_i(t) 
    \end{array} \right) = \left( \begin{array}{c:c}
        P_{\mathrm{\tilde{y}}_i \mathrm{\tilde{u}}_i}& P_{\mathrm{\tilde{y}}_i \mathrm{{u}}_i}\\ \hdashline 
         P_{\mathrm{{y}}_i \mathrm{\tilde{u}}_i}& P_{\mathrm{{y}}_i \mathrm{{u}}_i}
    \end{array} \right)  \left( \begin{array}{c}
         \tilde{u}_i(t) \\  {u}_i(t) 
    \end{array} \right),
    \label{LFR_SECTIONIII}
\end{equation}

\noindent where $u_i(t), y_i(t)\in \mathbb{R}^{n_{_{\mathrm{RB}}}}$ correspond to the filter inputs and outputs and  $\Tilde{u}_i(t), \Tilde{y}_i(t) \in \mathbb{R}^{n_{\mathrm{par}}^i}$ are latent variables, describing the interconnections between the parameter blocks and the filter blocks, where $n_{\mathrm{par}}^i$ describes the number of filter parameters per specific filter $i \in [1 \ n_\mathrm{F}]$. $n_\mathrm{F}$ represents the total number of filters considered during controller design.  By reformulating individual controller components into LFR representation, a modular controller design approach is introduced, facilitating the derivation of diverse controller structures such as: (i) cascade LFR interconnection, (ii) parallel LFR interconnection and (iii) even hierarchical feedback loops. Integration of these LFR representations yields a new LFR, with corresponding controller parameters consolidated into a diagonal matrix. Consequently, the closed-loop interconnection scheme, as depicted in Figure \ref{fig:SectionIII_ControllerStructure}, is reformulated into a generalized plant description ${\mathfrak{P}}$, where LFR representations of controller filters are absorbed. As a result, the resulting diagonal generalized controller block $\mathfrak{K}=\mathrm{diag}(\lbrace \Phi_i \rbrace_{i=1}^{n_\mathrm{F}})$ exclusively encompasses controller parameters subject to optimization.
%It is noteworthy to observe that through the extraction of the controller parameters, the resulting LFR, which is of form (\ref{LFR_SECTIONIII}), is LTI and exclusively comprises integrator connections. By reformulating individual controller components into LFR form, a modular structured controller design approach is introduced, wherein all feasible controller structures can be derived from two interconnection types: (i) cascade LFR interconnection and (ii) parallel LFR interconnection. Note that interconnection of these LFR representations results in a new LFR, where the corresponding controller parameters are stacked into a diagonal parameter matrix. In this context, the closed-loop interconnection scheme, illustrated by Figure \ref{fig:SectionIII_ControllerStructure}, can be reformulated into a generalized plant description, where the LFR representations of the controller filters are absorbed into the generalized plant. Consequently, the resulting diagonal controller block $K=\mathrm{diag}(\lbrace \Phi_i \rbrace_{i=1}^{n_\mathrm{F}})$ exclusively encompasses the controller parameters subject to optimization, where $n_{\mathrm{F}}$ corresponds to the number of filters considered during controller design.
%An additional benefit of the proposed approach is that (position dependent) static decoupling matrices can be incorporated in the structured controller, allowing for full block MIMO controller synthesis.
Moreover, the presented structured controller parameterization affords both modularity and scalability in the design of controllers across various structural configurations. This proposed controller parameterization is referenced as Contribution (C1) in this paper.

\section{Auto-tuning}
\label{SectionIV_Optimization}

This Section presents a novel frequency domain-based auto-tuning approach for structured LPV MIMO controllers based on the control interconnection illustrated in Figure \ref{fig:SectionIII_ControllerStructure}. Consider the non-convex objective function:
\vspace{-.2cm}
 \begin{equation}
\min_{\mathfrak{{{K}}}} \|\lbrace M_i\rbrace_{i=1}^n \|_{\mathcal{L}_{\infty}} + \Lambda_{\mathrm{stab}},
    \label{SectionII_CostFunction}
\end{equation}

\vspace{-.2cm}
\noindent 
where $\lbrace M_i \rbrace_{i=1}^n$ denotes a collection of weighted generalized plants, and $\Lambda_{\mathrm{stab}}$ represents a stability constraint. Notably, the objective function seeks to minimize the $\mathcal{L}_\infty$-norm of the set of weighted plants through optimization of the parameter matrix $\mathfrak{K}$, while penalizing closed-loop stability through $\Lambda_{\mathrm{stab}}$. Note that in case closed-loop stability is ensured, the $\mathcal{L}_\infty$-norm corresponds to the $\mathcal{H}_\infty$-norm, i.e., local $\mathcal{L}_2$-gain.

\subsection{Stability Analysis}
In this Subsection, an easy to check stability verification approach is presented which solely relies on lFRFs, and it is based on the principles of the Nyquist theorem, see \cite{skogestad2007multivariable}.  Closed-loop stability is assessed through the set of local MIMO loop transfers:
\vspace{-.2cm}
\begin{equation}
\lbrace L_i\rbrace_{i=1}^n =\lbrace \tilde{\mathcal{P}}_i{\mathcal{K}}_i\rbrace_{i=1}^n   .
\end{equation}

\vspace{-.2cm}
\noindent The generalized Nyquist theorem employs Cauchy's argument principle as a pivotal mechanism to ascertain the count of closed-loop poles situated within the ${D}$-contour, which delineates the right half of the complex plane,  see \cite{Oomen}. It is important to note that in the context of MIMO systems, poles positioned along the imaginary axis are encompassed within the ${D}$-contour. Let $P_{\mathrm{ol}}$ denote the number of open-loop poles of $\lbrace L_i \rbrace_{i=1}^n$ that reside within the $D$-contour. Then, the system is closed-loop stable if and only if the image of $\lbrace\mathrm{det}\left(I+ L_i(j\omega) \right)\rbrace_{i=1}^n$ makes $P_{\mathrm{ol}}$ counterclockwise encirclements of the origin and does not pass through the origin as $\omega$ traverses the $D$-contour in clockwise direction. Note that the image of $\lbrace\mathrm{det}\left(I+ L_i(j\omega) \right)\rbrace_{i=1}^n$ can be constructed through lFRFs at the observed frequency points, allowing for the assessment of local closed-loop stability. 
However, analysis of the encirclements made by  $\lbrace\mathrm{det}\left(I+L_i(j\omega) \right)\rbrace_{i=1}^n$ presents a significant computational challenge. Specifically, the presence of the large number of integrators in MIMO systems, originating from RB modes and integral action of the structured feedback controller, results in the low-frequency behavior of the image exhibiting computationally unreliable characteristics, as the amplitude tends toward infinity. As a solution to this issue, a novel extension of the factorized Nyquist check is presented which decomposes the contour into a more computationally attractive alternative. Note that this is a generalization of the factorized Nyquist theorem, see \cite{skogestad2007multivariable}, towards full block MIMO controllers. For a given local dynamics $\tilde{\mathcal{P}}_i$, consider the local MIMO interaction term $E_{i}$:
\vspace{-.2cm}
\begin{equation}
  E_{i}  = \left(\tilde{\mathcal{P}}_{i} {\mathcal{K}}_{i} -\hat{\mathcal{P}}_{i}\hat{\mathcal{K}}_{i} \right)\left(\hat{\mathcal{P}}_{i} \hat{ \mathcal{{K}}}_{i}\right)^{-1},
\end{equation}

\vspace{-.2cm}
\noindent where $\mathcal{K}$ is a full-block MIMO controller and:
\vspace{-.1cm}
\begin{equation}
 \hat{{\mathcal{P}}}_{i}= \mathrm{diag}( \lbrace \tilde{\mathcal{P}}_{i}^{jj} \rbrace_{j=1}^{n_{_{\mathrm{RB}}}} ), \qquad
 \hat{\mathcal{K}}_{i} = \mathrm{diag}( \lbrace {\mathcal{K}}_{i}^{jj} \rbrace_{j=1}^{n_{\mathrm{RB}}} ), 
\end{equation}

\vspace{-.1cm}
\noindent which allows for decomposition of the Nyquist criterion as:
\vspace{-.1cm}
\begin{equation}
  I+\tilde{\mathcal{P}}_{i}\mathcal{K}_{i}   = \left(I+E_{i}{\mathcal{T}}_{i}\right )(  I+{\hat{\mathcal{P}}}_{i}\hat {\mathcal{K}}_{i}) ,
    \label{Extended_Fact_Nyq}
\end{equation}

\vspace{-.2cm}
\noindent where ${\mathcal{T}}_{i}  = \hat{\mathcal{P}}_{i}\hat{\mathcal{K}}_{i}(I+\hat{\mathcal{P}}_{i}\hat{\mathcal{K}}_{i})^{-1}  $. This reformulation results in a decomposition of the stability assessment as:
\vspace{-.2cm}
\begin{equation}
\mathrm{det}\left(I+L_i \right)=
  \mathrm{det}\left( I+E_i {\mathcal{T}}_i \right) \cdot \prod_{j=1}^{n_{\mathrm{RB}}} \left(1+ \tilde{\mathcal{P}}_i^{jj} {\mathcal{K}}_i^{jj}  \right). 
   \label{DecomposedStabilityAssessment}
\end{equation}

\vspace{-.2cm}
\noindent 
From (\ref{DecomposedStabilityAssessment}) it is observed that assessing the stability of $\mathrm{det}(I+L_i)$ is decomposed into $n_{_{\mathrm{RB}}}+1$ encirclement checks, whereby the parts containing integrators are decoupled into SISO checks, ensuring the computational feasibility of the algorithm.  The extended stability check for full block MIMO controllers based on lFRF data corresponds to Contribution (C2) of this paper.

%Moreover, let $P_{\mathrm{ol}}$ be the number of open-loop poles of $\lbrace L(j\omega) \rbrace_{i=1}^n$ that lie in the $D$-contour. Then the closed-loop system is stable if the image of $\mathrm{det}\left(I+\lbrace L(j\omega)\rbrace_{i=1}^n \right)$ makes: (i) $P_{\mathrm{ol}}$ counterclockwise encirclements of the origin, and (ii) does not pass the origin, as $\omega$ traverses the $D$-contour.   

\subsection{Performance Shaping}
\label{PerformanceShaping}

In this Subsection, a norm-based performance optimization approach is presented, with particular reference to the set of weighted closed-loop lFRFs $\lbrace M_i \rbrace_{i=1}^n$ in (\ref{SectionII_CostFunction}). %Consider the control interconnection illustrated by Figure \ref{fig:SectionIII_ControllerStructure}, where performance variables $w_1,w_2,z_1,z_2 \in \mathbb{R}^{n_{\mathrm{RB}}\times n_{\mathrm{RB}}}$ are shaped to attain desired closed-loop performance. 
The non-convex nature of the cost function permits decomposition of the commonly employed 4-block shaping configuration, see \cite{van2002multivariable}, into four distinct shaping problems sharing common controller parameters. In this context, by selecting the performance channels as $w(t) = (\eta(t) \ d(t) \ \eta(t) \ d(t))^\top$ and $z(t) = (e(t) \ e(t) \ u(t) \ u(t))^\top$, as shown in Figure \ref{fig:SectionIII_ControllerStructure}, the following set of weighted closed-loop dynamics is obtained:
\vspace{-.2cm}
\begin{equation}
\resizebox{.89\hsize}{!}{$
    \lbrace M\rbrace_{i=1}^n = \left\lbrace W_\mathrm{z} \cdot   \mathrm{diag}\left( \mathcal{S}_i,{\mathcal{K}}_i \mathcal{S}_i, \mathcal{S}_i{\tilde{\mathcal{P}}}_i,{\mathcal{K}}_i \mathcal{S}_i{\tilde{\mathcal{P}}}_i\right)\right \rbrace_{i=1}^n  ,$}
    \label{FullWeightedPlant} 
\end{equation}

\vspace{-.1cm}
\noindent where $W_\mathrm{z} = \mathrm{diag}(W_\mathrm{z}^\mathcal{S},W_\mathrm{z}^\mathcal{KS},W_\mathrm{z}^{\mathcal{S}\tilde{\mathcal{P}}},W_\mathrm{z}^{\mathcal{KS}\tilde{\mathcal{P}}})$ corresponds to an output shaping filter and $\lbrace\mathcal{S}_i \rbrace_{i=1}^n = \lbrace(I+{\tilde{\mathcal{P}}}_i{\mathcal{K}}_i )^{-1}\rbrace_{i=1}^n$. Note that the partitioning of the shaping loops yields increased flexibility in the design of $W_\mathrm{z}$, due to the decoupling of the closed-loop sensitivities. To capitalize on the non-convex characteristics of (\ref{SectionII_CostFunction}), the shaping filters are designed as piece-wise affine functions of frequency, thereby affording increased design flexibility compared to conventional frequency-domain based shaping filters. In this context, the sensitivity shaping filter $W_\mathrm{z}^{\mathcal{S}}$ is designed to be of form:
\begin{align}
W_\mathrm{z}^{\mathcal{S}}=
K_s \cdot \mathrm{diag}\left(
    \begin{cases}
            \frac{(\alpha^{-1} \omega_{\mathrm{bw}}^i)^3}{\omega^3 },  &\mathrm{if} \ \ \omega \leq \frac{\omega_{\mathrm{bw}}^i}{\alpha} \\
            1,  &\mathrm{else}
    \end{cases} \right),
    \label{ShapingFilterSensitivity}
\end{align}

\noindent where $K_s$ is set to 0.5$I$ to impose a 6dB upper-bound on the sensitivity, $\omega_{\mathrm{bw}}^i$ corresponds to the target bandwidth of the $i$-th RB channel and $\alpha > 1$ is a tuning parameter that is used for sharpening of the sensitivity constraints. Similarly, the shaping filter for the complementary sensitivity $W_\mathrm{z}^{\mathcal{KS}\Tilde{\mathcal{P}}}$ is defined as:
\begin{align}
W_\mathrm{z}^{\mathcal{KS}\Tilde{\mathcal{P}}}=
K_{r} \cdot\mathrm{diag}\left(
    \begin{cases}
               \frac{\omega}{\alpha \omega_{\mathrm{bw}}^i},  &\mathrm{if} \ \ \omega \geq \omega_{\mathrm{bw}}^i \alpha\\
                1,  &\mathrm{else} 
    \end{cases} \right),
        \label{ShapingFilterKSensitivity}
\end{align}

\noindent where $K_r$ is typically chosen as 0.5$I$ to place a 6 dB upper-bound on the complementary sensitivity.
The control sensitivity shaping filter $W_\mathrm{z}^{\mathcal{KS}}$ is designed of the same structure as (\ref{ShapingFilterKSensitivity}). The design of $K_r$ involves channel scaling, accomplished by setting $K_r = \mathrm{diag}(\mathrm{abs}((\mathcal{P}^{ii}(j\omega_{bw}))))$ for all $i \in [1 \ n_{_{{\mathrm{RB}}}}]$. This selection corresponds to the worst-case gain concerning the modulus margin at the target bandwidth $\omega_{\mathrm{bw}}^j$, with an additional 6 dB margin applied on top of the filter. In a similar manner, the scaling of the process sensitivity is achieved by $K_p = \mathrm{diag}(\mathrm{abs}((\mathcal{P}^{ii}(j\omega_{bw})))^{-1})$, where the shaping filter for the process sensitivity $W_\mathrm{z}^{\mathcal{S}\mathcal{P}}$ corresponds to:
\begin{align}
W_\mathrm{z}^{\mathcal{S}\Tilde{\mathcal{P}}}= K_p \cdot \mathrm{diag}\left(
    \begin{cases}
            \frac{(\alpha^{-1} \omega_{\mathrm{bw}}^j)}{\omega },  &\mathrm{if} \ \ \omega \leq \frac{\omega_{\mathrm{bw}}^j}{\alpha} \\
             \frac{\omega}{\alpha \omega_{\mathrm{bw}}^j},  &\mathrm{if} \ \ \omega \geq \omega_{\mathrm{bw}}^j \alpha \\
            1,  &\mathrm{else}
    \end{cases} \right).
    \label{shapingfilterprocesssensi}
\end{align}

\subsection{Solving the optimization problem}

To facilitate the auto-tuning of the structured LPV feedback controller, first, a desired feedback control structure must be specified according to the controller parameterization presented in Section \ref{SectionIII_Approach}. Next, a set of weighted generalized plants is constructed by collapsing both the LFRs of the structured controller, and the desired shaping filters, e.g. (\ref{ShapingFilterSensitivity}), (\ref{ShapingFilterKSensitivity}) and  (\ref{shapingfilterprocesssensi}), into a weighted generalized plant representation. The resulting generalized controller $\mathfrak{K}$ takes the form of a diagonal gain matrix comprising the controller parameters to be optimized. To optimize the non-convex cost function (\ref{SectionII_CostFunction}), we propose a two-step optimization strategy to converge towards a globally sub-optimal solution, aiming to minimize the $\mathcal{H}_\infty$-norm of the weighted plants $\lbrace M_i\rbrace_{i=1}^n $. This objective is achieved by penalizing closed-loop stability through the constraint $\Lambda_{\mathrm{stab}}$, which is evaluated using (\ref{DecomposedStabilityAssessment}). If the closed-loop is stable, $\Lambda_{\mathrm{stab}}$ is excluded from the cost function; however, in the case of instability, $\Lambda_{\mathrm{stab}}$ is assigned a high-cost penalty. The optimization process is initiated with \emph{particle swarm optimization} (PSO), a gradient-free optimization algorithm well-regarded for its ability to explore the global search space for optimal parameters, see \cite{eberhart1995particle}. Nevertheless, due to the random exploration of PSO, it can reach the neighborhood of the global optimum but its asymptotic convergence is slow and can only be guaranteed in a probabilistic sense. Consequently, a secondary optimization step is introduced, employing gradient-descent methodologies, particularly the BFGS optimization technique, see \cite{international1990bfgs}. This step is initialized with the PSO-based solution, thereby improving cost optimization by convergence towards a minimum, therefore satisfying (R1) and (R2).

\subsection{Controller implementation}
In this subsection, a novel discretization approach is introduced that maintains the parameterization of CT controllers, preserving their physical interpretation in the resulting \emph{discrete-time} (DT) domain. Leveraging the modularity of the structured controller design outlined in Section \ref{SectionIII_Approach},
the LFR representation of the controller is discretized. The resulting DT controller is obtained by collapsing back the CT parameters obtained from auto-tuning. Consider the CT state-space representation \TR{of the transfer function} (\ref{LFR_SECTIONIII}):
\begin{equation}
I s^{-1} \star \left(\begin{array}{cc}
A_i & B_i \\ C_i & D_i
\end{array} \right),
    \label{DTSYSTEM}
\end{equation}
\noindent where \TR{$s\in\mathbb{C}$ is the complex frequency} and $\star$ is the \emph{star product}. Note that in the time-domain, ${s}^{-1}$ is replaced by an integrator, \TR{expressed as the operator $\delta^{-1}$}. Inspired by \cite{whitbeck1978digital}, to maintain the CT parameterization of the controller, our objective is to articulate the discretization utilizing a time-domain operator, as opposed to employing matrix operations to transform the state-space realization to the $z$-domain. Consider the Tustin discretization, which \TR{corresponds to} %relates to the Laplace \TR{domain} as:
%In \cite{whitbeck1978digital}, a frequency domain discretization technique has been introduced capable of retaining the CT parameterization through a so called \emph{w'}-domain transformation:
\begin{equation}
        s= \frac{2}{T_\mathrm{s}} \frac{z-1}{z+1} \quad \Rightarrow \quad \delta = \frac{2}{T_\mathrm{s}} \frac{q-1}{q+1},
        \label{W_Domain_Freq_Operator}
\end{equation}
\noindent  where $T_\mathrm{s}$ is the sampling time, $q$ denotes the \TR{forward} time-shift operator, \TR{and $\delta$ is the differentiation operator. Note that here the equality corresponds to substitution of $s$ and $\delta$. Then, we can provide a filter representation of the $\delta$ provided mapping} 
\begin{equation}
    \TR{{\tt y}(k) =   \frac{2}{T_\mathrm{s}} \frac{q-1}{q+1} {\tt u}(k),}
            \label{W_Domain_Freq_Operator2}
\end{equation}
%In this context, we introduce an auxiliary variable $\lambda(k)$ to decouple the input and output delays: 
%\begin{equation}
%    \frac{\TR{\tt y}(k)}{\TR{\tt u}(k)} =  \frac{\TR{\tt y}(k)}{\lambda(k)} \frac{\lambda(k)}{\TR{\tt u}(k)}  = \frac{2}{T_\mathrm{s}} \frac{q-1}{q+1},
%            \label{W_Domain_Freq_Operator2}
%\end{equation}
\noindent
where  $k \in \TR{\mathbb{Z}_{0}^+}$ is the discrete time and \TR{${\tt u}(k),{\tt y}(k)$} correspond to the input and output of the operator,  \TR{in the form of} %Using the auxiliary state $\lambda(k)$, we can introduce a time-domain equivalent of \eqref{W_Domain_Freq_Operator} in terms of an $r$-operator:
\begin{equation}
    r(q) := \begin{cases}
        \begin{split}
            \lambda(k+1) &= -\lambda(k) + T_\mathrm{s}^{-1}\TR{\tt u}(k),\\
            \TR{\tt y}(k) &= -4 \lambda(k) + 2 T_\mathrm{s}^{-1}\TR{\tt u}(k),
        \end{split}
    \end{cases}
\end{equation}
\TR{which is a time-domain equivalent of \eqref{W_Domain_Freq_Operator} in terms of an $r$-operator and $\lambda(k)$ as an auxiliary state.}
\TR{The integrator}, i.e., $\TR{\delta}^{-1}$, can be expressed by inverting the $r$-operator in the time-domain as:
\begin{equation}
    r^{-1}(q) := \begin{cases}
        \begin{split}
            \lambda(k+1) &= \lambda(k) + \frac{1}{2}\TR{\tt y}(k),\\
            \TR{\tt u}(k) &= 2T_\mathrm{s}  \lambda(k) +  \frac{1}{2} T_\mathrm{s}   \TR{\tt y}(k).
        \end{split}
    \end{cases}
    \label{W_Domain_Freq_Operator3}
\end{equation}

\noindent
By substituting \eqref{W_Domain_Freq_Operator3} for $s^{-1}$, we obtain the Tustin-discretized representation of \eqref{DTSYSTEM}, while preserving the CT controller matrices:
\begin{equation}
I r^{-1}(q) \star \left(\begin{array}{cc}
A_i & B_i \\ C_i & D_i
\end{array} \right).
    \label{DTSYSTEM32}
\end{equation}

\noindent 
The interconnection of (\ref{DTSYSTEM32}) is well-posed if and only if $\mathrm{det}(I-A_i\frac{T_\mathrm{s}}{2})\neq 0$, a condition automatically satisfied due to the LFR form of the controller. Additionally, note that the $r^{-1}$-operator can be directly integrated into the optimization of the CT controller parameters using (\ref{SectionII_CostFunction}), allowing for direct DT controller synthesis through CT parameterization. However, this adjustment necessitates adaptation of the stability analysis, as encirclements must now be considered along the $C$-contour, see \cite{yeung1988reformulation}, and the considered lFRFs must be treated as discrete-time ones. The presented DT controller implementation corresponds to Contribution (C3) in the paper.
 
\section{Experimental Validation}
\label{SectionV_ExperimentalValidation}
\subsection{MMPA Prototype System}

%A MMPA system, of which a prototype is illustrated Figure \ref{fig:MMPA}, is a high-precision motion system which exhibits position dependent flexible dynamics due to position dependent actuation and sensing of the mover.
%Consider the MMPA prototype illustrated in Figure \ref{fig:MMPA}.

\begin{figure}[t]
    \centering
\includegraphics[height=3.8cm,width=.9\linewidth]{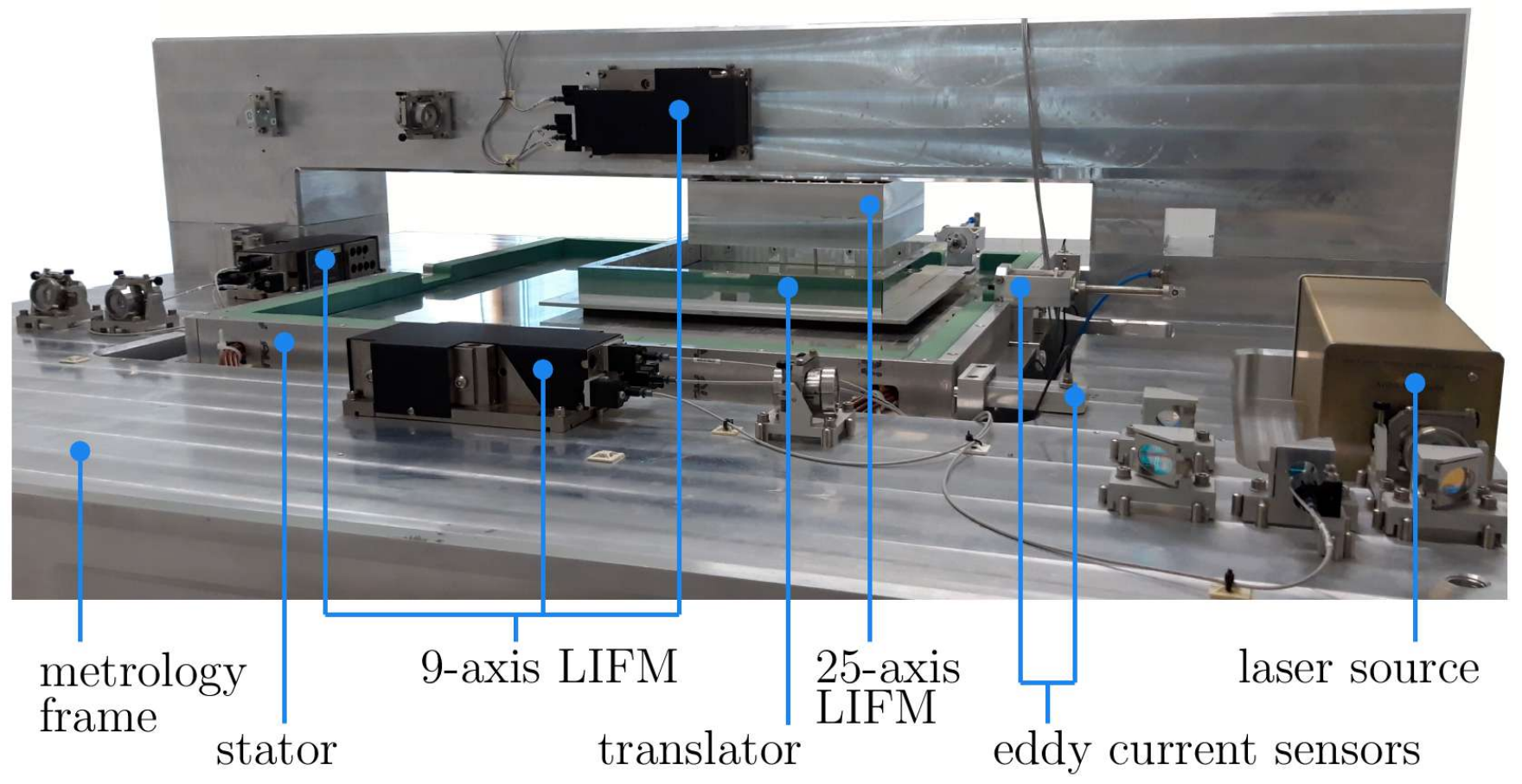}
    \caption{Photograph of a moving-magnet planar actuator system prototype.}
    \label{fig:MMPA}
    \vspace{-.5cm}
\end{figure}

A MMPA system, illustrated in Figure \ref{fig:MMPA}, displays position-dependent effects due relative actuation and sensing of the moving-body. Comprising three key components, this system includes a stator base with a double-layer coil array, a lightweight translator equipped with a Halbach array of 281 permanent magnets, and a metrology frame featuring 9 laser interferometers for precise displacement measurements of the translator. For a comprehensive overview of such a prototype, refer to \cite{proimadis2021active}.

%By replacing the complex variable $s$ by $r$, (\ref{DTSYSTEM}) is represented in the $w'$ domain as: 
%\begin{equation}
%\begin{split}
%\left(\begin{array}{c}
%r{x}_i(t) \\ \hdashline y_i(t) \\ z_i(t)
%\end{array} \right) = \left(\begin{array}{c:c}
%A_i &B_i  \\ \hdashline 
%C_i &D_i 
%\end{array} \right) \left(\begin{array}{c}
%x_i(t) \\ \hdashline u_i(t) \\ w_i(t)
%\end{array} \right),
%\end{split}
%    \label{DTSYSTEM1}
%\end{equation}

%The discrete-time implementation scheme, as illustrated in Figure \ref{fig:RoperatorImplementation}, reveals the presence of an algebraic loop arising from the feedback interconnection of $A_i$ with the $r^{-1}$ operator. Consequently, this interconnection is deemed well-posed if and only if $\mathrm{det}(I-A\frac{T_\mathrm{s}}{2}) \neq 0$. However, owing to the parameterization of the LFR controller as delineated in Equation (\ref{SectionIII_Sigma_1}), the generalized plants exclusively incorporate integrators. Thus, the integrators may be substituted with the $r^{-1}$ operator:
%\begin{equation}
%\left(\begin{array}{c}
%\xi(k+1) \\ x(k)
%\end{array} \right) = %\left(\begin{array}{cc}
%I&2I\\
%\frac{T_\mathrm{s}}{2}I&\frac{T_\mathrm{s}}{2}I
%\end{array} \right)\left(\begin{array}{c}
%\xi(k) \\ rx(k)
%\end{array} \right),
%\end{equation}

%\noindent 
%thereby satisfying the well-posedness condition. Moreover, discretization is achieved, while preserving the CT parameterization of the controller, thereby satisfying (R3). The presented $r^{-1}$ operator discretization corresponds to Contribution (C3) of the paper.

\vspace{-.05cm}
\subsection{Experimental Results}

To showcase the effectiveness of the structured feedback control auto-tuning approach, two types of MIMO controllers were synthesized employing the methodology detailed in Section \ref{SectionIV_Optimization}. Each controller configuration comprises a PI-controller described by (6), augmented by three lead filters of the form (7), and a notch filter of form (8). It is noteworthy that for the robust controller, the notch filter is designed invariant of position, while for the LPV controller, the notch coefficients are assumed to exhibit a first-order polynomial dependence on the scheduling vector, i.e., $\eta_x, \eta_y$. Synthesis of both controllers utilized 11 lFRFs of the MMPA prototype via the shaping approach outlined in Subsection \ref{PerformanceShaping}.
\begin{figure}[b]
\vspace{-.4cm}
    \centering  \includegraphics[trim={1.7cm 0cm 1.7cm 0cm}
    ,width=.91\linewidth]{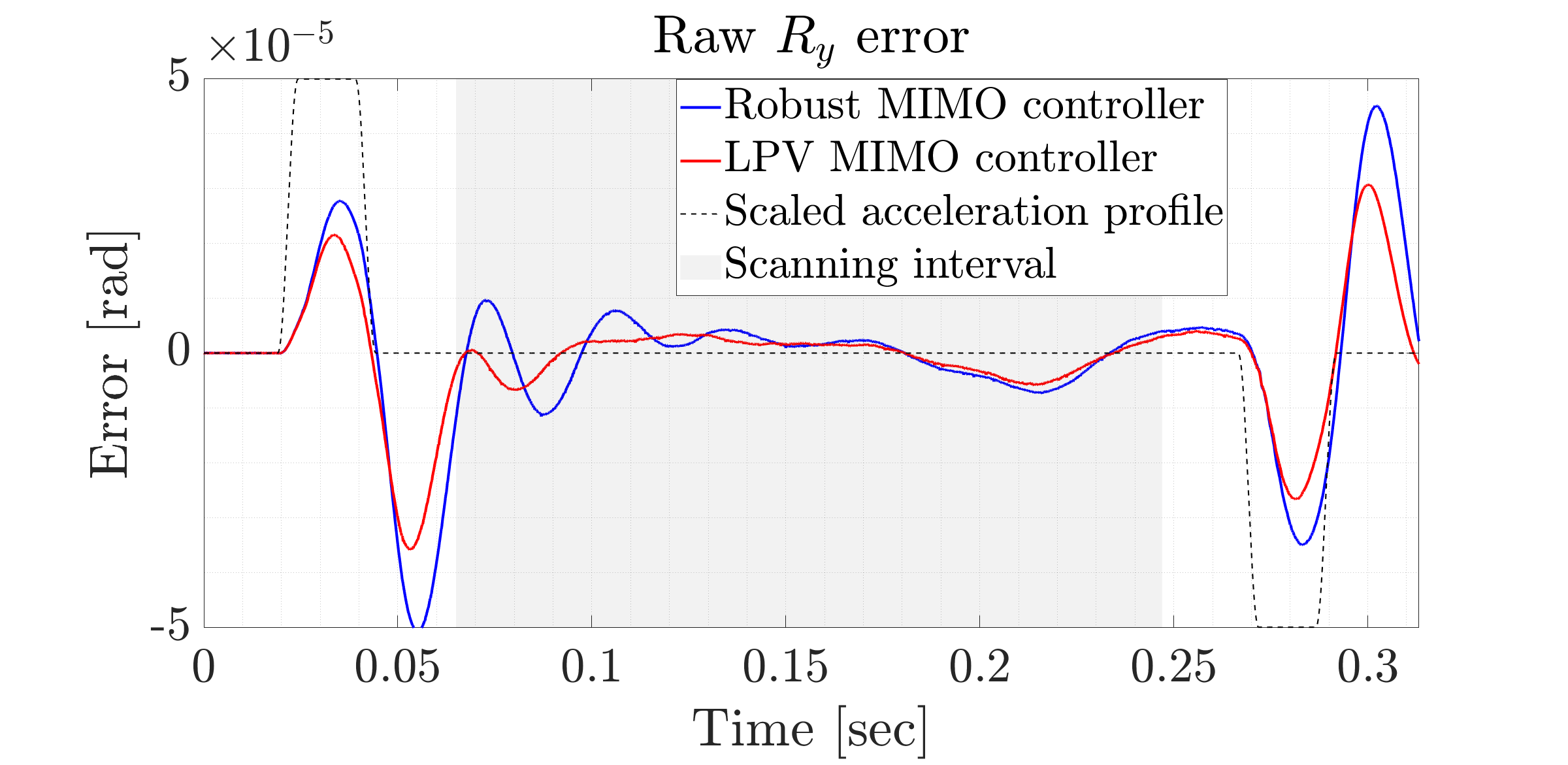}
    \caption{Position tracking error in $R_y$ direction during the constant velocity interval of the motion profile with: (\Large\textcolor{blue} {-}\footnotesize) Robust controller, (\Large\textcolor{red}{-}\footnotesize) LPV controller.}
    \label{fig:PositionTrackingError}
\end{figure}
To experimentally validate the efficacy of these controllers, lithographic scanning motions were performed, see  \cite{Butler}, in both the $x$ and $y$ directions simultaneously, employing a fourth-order motion profile, with $a_{\mathrm{max}}=10 \frac{m}{s^2}$, $v_{\mathrm{max}}= 0.2 \frac{m}{s}$, and maximum displacement in both $x$, $y$ direction of $0.05 m$.

The experimental results in Figure \ref{fig:PositionTrackingError}, obtained under identical conditions, compare the position tracking error in the $R_y$ direction for two controllers: the robust controller (blue graph) and the LPV controller (red graph). The $R_y$ axis was chosen due to its high sensitivity to position-dependent flexible dynamics, see Figure \ref{fig:SectionII_FigureI}. The LPV controller outperforms the robust controller, reducing the worst-case error during the scanning interval, i.e., constant velocity interval, from $11.55 \times 10^{-6}$ rad to $6.78 \times 10^{-6}$ rad, resulting in a relative improvement of 43.10\%. This is attributed to the position-dependent notch filter used in the LPV controller, allowing for increased rigid-body feedback control bandwidth.

\section{Conclusion}
\label{SectionVI_Conclusion}

This paper introduces a novel frequency domain auto-tuning technique for structured LPV MIMO controllers, relying solely on FRF data. The approach includes a new controller parameterization scheme, enabling modular structured controller synthesis. Experimental validation on an MMPA prototype demonstrates the method's effectiveness, with both robust and LPV MIMO controllers synthesized. Notably, the LPV controller achieved a 43.10\% relative performance improvement in the $R_y$-direction compared to the robust design.
\addtolength{\textheight}{-12cm}   % This command serves to balance the column lengths

%%%%%%%%%%%%%%%%%%%%%%%%%%%%%%%%%%%%%%%%%%%%%%%%%%%%%%%%%%%%%%%%%%%%%%%%%%%%%%%%

%%%%%%%%%%%%%%%%%%%%%%%%%%%%%%%%%%%%%%%%%%%%%%%%%%%%%%%%%%%%%%%%%%%%%%%%%%%%%%%%

%%%%%%%%%%%%%%%%%%%%%%%%%%%%%%%%%%%%%%%%%%%%%%%%%%%%%%%%%%%%%%%%%%%%%%%%%%%%%%%%
%\section*{APPENDIX}

\bibliographystyle{ieeetr}   
\bibliography{MyBib}

\end{document}